# Impacts of Electric Vehicle Charging Regimes and Infrastructure Deployments on System Performance: An Agent-Based Study

1st Jiahua Hu
*Department of Civil and Environmental Engineering*
*Monash University*
Melbourne, Australia
Jiahua.hu@monash.edu

2nd Hai L. Vu
*Department of Civil and Environmental Engineering*
*Monash University*
Melbourne, Australia
Hai.Vu@monash.edu

3rd Wynita Griggs
*Department of Electrical and Computer Systems Engineering*
*Monash University*
Melbourne, Australia
Wynita.Griggs@monash.edu

4th Hao Wang
*Department of Data Science & AI*
*Monash University*
Melbourne, Australia
Hao.Wang2@monash.ed

***Abstract*—The rapid growth of electric vehicles (EVs) requires more effective charging infrastructure planning. Infrastructure layout not only determines deployment cost, but also reshapes charging behavior and influences overall system performance. In addition, destination charging and en-route charging represent distinct charging regimes associated with different power requirements, which may lead to substantially different infrastructure deployment outcomes. This study applies an agent-based modeling framework to generate trajectory-level latent public charging demand under three charging regimes based on a synthetic representation of the Melbourne (Australia) metropolitan area. Two deployment strategies, an optimization-based approach and a utilization-refined approach, are evaluated across different infrastructure layouts. Results show that utilization-refined deployments reduce total system cost, accounting for both infrastructure deployment cost and user generalized charging cost, with the most significant improvement observed under the combined charging regime. In particular, a more effective allocation of AC slow chargers reshapes destination charging behavior, which in turn reduces unnecessary reliance on en-route charging and lowers detour costs associated with en-route charging. This interaction highlights the behavioral linkage between destination and en-route charging regimes and demonstrates the importance of accounting for user response and multiple charging regimes in charging infrastructure planning.**

## I. Introduction

The global electric vehicle (EV) fleet is projected to grow rapidly by 2030, accompanied by substantial expansion of public and fast-charging infrastructure [1]. In Australia, although more than 1,300 fast and ultra-fast charging sites were operational by mid-2025, around 45 EVs still share each public charging plug, indicating persistent utilization pressure [2]. These developments highlight the need for effective charging infrastructure planning under increasing demand. Early planning studies typically modeled charging demand as exogenous inputs based on aggregate spatial or flow representations [3], [4], which, while computationally tractable, abstract from individual activity schedules, route choices, and energy-constrained decision processes, limiting their ability to capture behavioral adaptation to infrastructure availability and network conditions [5], [6].

To address this limitation, agent-based transport simulation frameworks have been adopted to generate charging demand endogenously from individual activity-travel pattern. By explicitly modeling activity scheduling, route selection, and vehicle's battery State of Charge (SoC) dynamics, these models allow charging demand to emerge from agents' interactions with the mobility and charging environment s [7]. However, due to their computational complexity, many studies employ simulation outputs in a sequential manner, using simulated demand as fixed input to subsequent optimization models [8], [9]. In such frameworks, infrastructure design is evaluated against simulated demand, but behavioral responses to alternative infrastructure layouts are not systematically studied.

In practice, charging infrastructure affects accessibility, detour cost, and waiting conditions, which in turn reshapes charging decisions and spatial demand. Simulation-based optimization approaches have therefore emerged to better account for this interaction [10], [11]. Nevertheless, most existing studies focus on identifying an optimal spatial deployment at the zone or station level, while charger composition and the dynamic interaction between behavioral response and infrastructure configuration remain insufficiently examined [12].

Meanwhile, the growing deployment of high-power fast charging has increased the feasibility of corridor-oriented en-route charging, while activity-based destination charging continues to dominate daily urban travel. These regimes impose fundamentally different spatial and power requirements, making coordinated infrastructure design increasingly important [13]. However, most existing studies focus on a single charging regime, either en-route or destination charging. Bi et al. distinguish mandatory and convenience charging but still within destination-based settings [14]. Although integrated planning frameworks for fast and slow charging facilities have been proposed, they typically rely on aggregate demand modeling and do not explicitly represent en-route and destination charging as behaviorally distinct processes [15]. Consequently, limited attention has been given to systematically comparing how alternative charging regimes lead to different infrastructure layouts and system-level outcomes under a unified modeling framework [16].

To address these gaps, this study develops an integrated modeling framework that explicitly links behaviorally generated charging demand with infrastructure design and evaluation. Using a synthetic population representing the Melbourne metropolitan area within an agent-based modeling environment, trajectory-level latent charging demand is generated under three charging regimes: destination charging, en-route charging, and a combined regime. Based on these demands, we solve a capacitated maximal covering location problem (CMCLP) following by a utilization-refined deployment which enables a systematic comparison across infrastructure layouts and charging regimes. By evaluating system-level performance and behavioral responses under a unified modeling framework, this study provides a structured assessment of the interaction between charging infrastructure configuration and EV charging behavior.

## II. METHODOLOGY

Fig. 1 presents the proposed methodological framework of this study. We develop a two-stage deployment generation and refinement process to obtain two distinct charging infrastructure layouts.

In the first stage, latent public charging demand is generated using BEAM (Behavior, Energy, Autonomy, and Mobility), a large-scale agent-based transportation simulation built upon MATSim [17], under predefined charging assumptions. Residential charging is assigned to a subset of EV households, while baseline public charging assumes sufficient charger availability at each Traffic Analysis Zone (TAZ) to allow immediate local charging when needed. The resulting latent charging demand is used as an input to a CMCLP, yielding an initial, demand-driven charging infrastructure deployment.

In the second stage, the CMCLP-based deployment is re-evaluated through agent-based simulation to capture charging station utilization arising from endogenous agent charging decisions. Utilization metrics, derived from delivered energy and theoretical charging capacity, are then used to apply a set of utilization-driven rules that adjust charger capacities at existing stations. This refinement process is repeated via iterations until the charging infrastructure configuration remains unchanged, resulting in a behavior anticipated deployment.

Different charging regimes generate distinct latent charging demand patterns in the first stage. Each regime therefore produces two corresponding deployments, a CMCLP-based deployment and a utilization-refined deployment, yielding six deployment scenarios to be evaluated in terms of user behavior and system-level performance.

### A. *Agent-Based Charging Decision Logic in BEAM*

BEAM agent-based simulation is designed to jointly represent travel behavior and vehicle energy dynamics [18]. By explicitly modeling vehicle energy consumption, heterogeneous powertrains, parking supply, and refueling infrastructure, BEAM enables the integrated simulation of mobility and charging decisions in urban transportation systems [19].

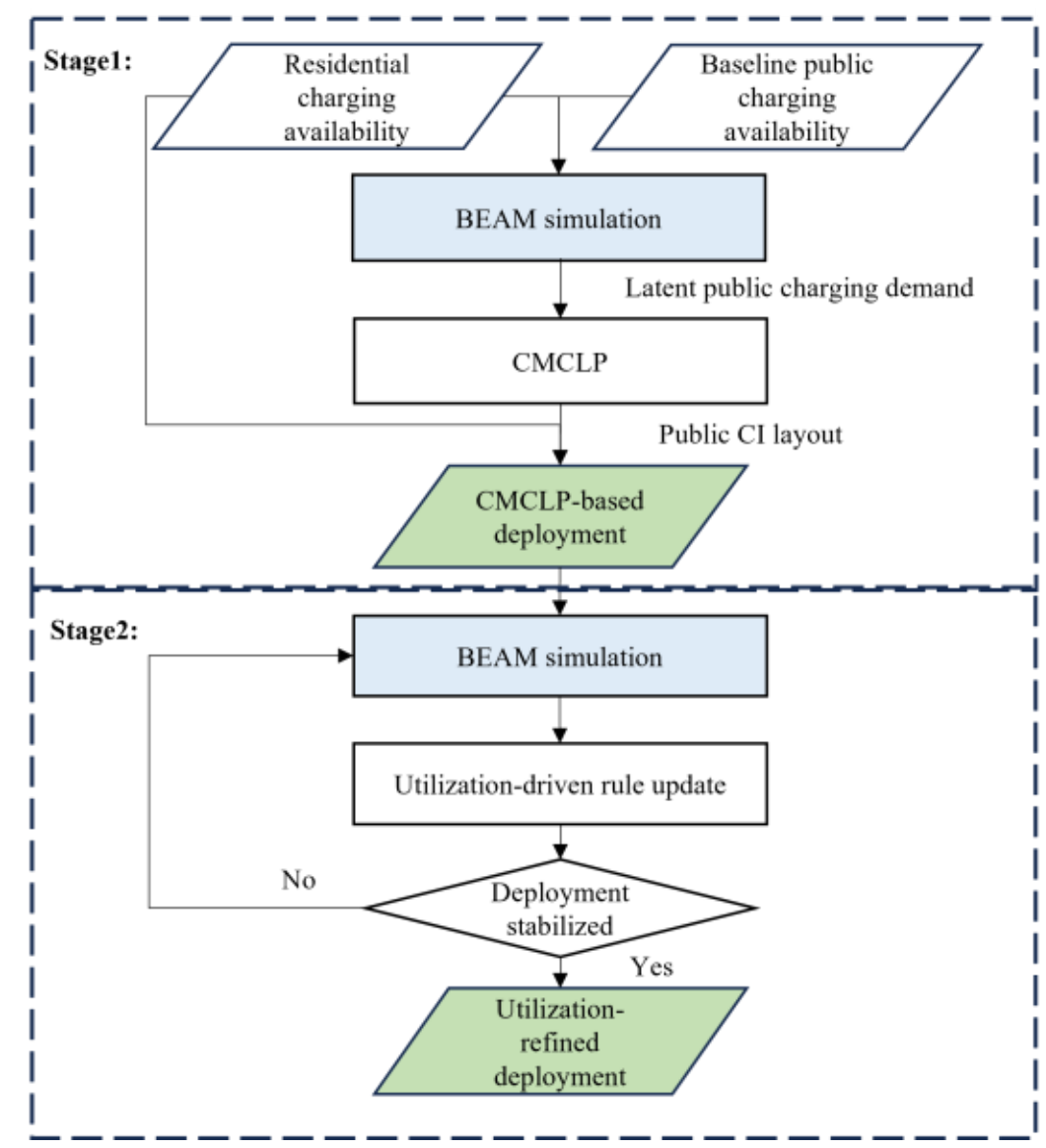

Fig. 1. Overview of the Two-Stage Deployment Generation Framework

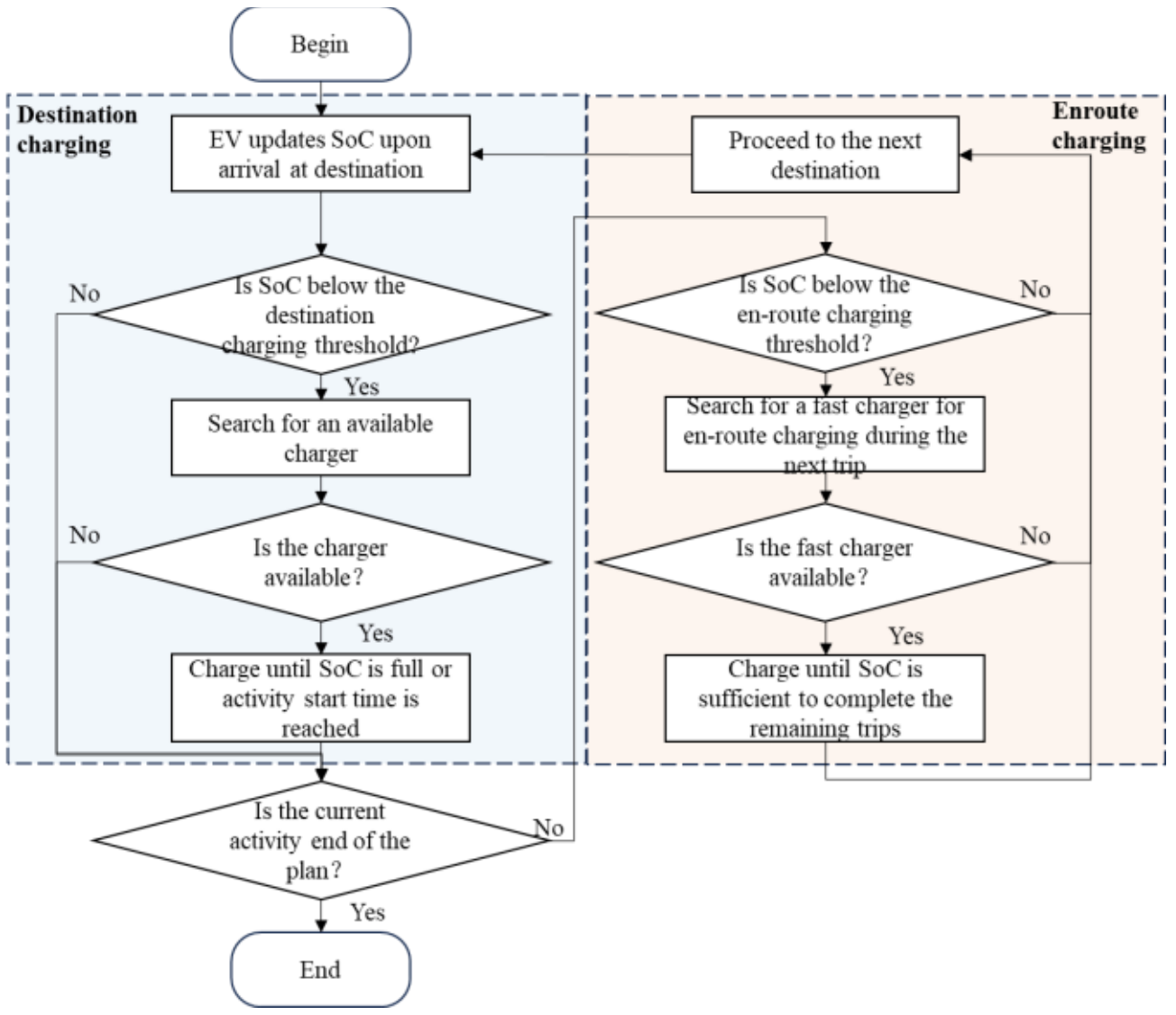

Fig. 2. Charging Behavior with Two Charging Regimes in BEAM

In BEAM, vehicle agents follow predefined daily activity plans and make charging decisions endogenously during simulation based on EV battery's SoC, activity schedules, and charging availability. Fig. 2 summarizes the charging decision logic, which distinguishes between destination charging and en-route charging.

$$P_{\text{dest}}^{(z)}(\text{SoC}) = \begin{cases} 1, & \text{SoC} \leq \underline{\theta}_z, \\ \frac{\bar{\theta}_z - \text{SoC}}{\bar{\theta}_z - \underline{\theta}_z}, & \underline{\theta}_z < \text{SoC} < \bar{\theta}_z, \\ 0, & \text{SoC} \geq \bar{\theta}_z, \end{cases} \tag{1}$$

Destination charging is evaluated upon arrival at activity locations. As shown in (1), the probability of triggering charging is determined by the current SoC relative to predefined threshold parameters, which differ between public and residential charging contexts. When charging is triggered, the agent searches for an available charger in the vicinity of the destination and commences charging subject to charger availability and activity duration constraints. En-route charging is evaluated prior to departure from an activity location. Before initiating the next trip, the agent assesses whether the current SoC is sufficient to complete the remaining planned trips. If the SoC is deemed insufficient, an additional detour trip is generated toward an available fast charger along or near the upcoming route segment. The agent then charges until sufficient energy is restored to continue the planned itinerary.

Through this mechanism, public charging demand in BEAM emerges from agents' travel patterns, activity schedules, and charging opportunities, rather than being imposed exogenously. The resulting latent charging demand provides the basis for the infrastructure deployment analysis conducted in this study.

### B. *Charging Infrastructure Deployment via CMCLP*

The deployment problem is formulated as a CMCLP on time-expanded charging demand. Time is discretized into 30-minute bins, and each charging event $k$ with required charger type $p_k$ is represented by time-indexed demand items $(k, b)$ over its occupied time bins. Candidate station locations $i \in I$ correspond to latent charging occurrence points extracted from simulation. Decision variables include station selection $x_i$, charger installation $n_{i,p}$, assignment variables $a_{i,kb}$, and coverage indicators $c_{kb}$. Let $\mathcal{K}$ denote the set of time-expanded demand items $(k, b)$, and $\mathcal{B}$ the set of discrete time bins. The objective maximizes the total number of covered time-expanded charging demand:

$$\max \sum_{(k,b)\in\mathcal{K}} c_{kb} \tag{2}$$

Each demand item can either be assigned to exactly one feasible station or remain uncovered:

$$\sum_{i\in\mathcal{N}(k)} a_{i,kb} = c_{kb}, \forall (k,b) \in \mathcal{K} \tag{3}$$

where $\mathcal{N}(k)$ denotes the set of candidate stations within a predefined service radius of 1 km from demand location $k$ [7].

Assignments $a_{i,kb}$ are restricted to stations that are selected, as indicated by $x_i$.

$$a_{i,kb} \le x_i, \forall i \in I, \forall (k,b) \in \mathcal{K} \tag{4}$$

Concurrent assignments of power type $p$ at station $i$ in time bin $b$ are limited by the installed charger capacity $n_{i,p}$.

$$\sum_{\substack{(k,b)\in\mathcal{K} \\ p_k = p}} a_{i,kb} \le n_{i,p}, \forall i \in I, \forall p \in \mathcal{P}, \forall b \in \mathcal{B} \tag{5}$$

The total investment, including charger-level and site-level costs, is constrained by the available budget $B$.

$$\sum_{i\in I}\sum_{p\in\mathcal{P}} n_{i,p}\left(c_p^{cap} + c_p^{op}\right) + \sum_{i\in I}\left(c_{AC}^{site}\left(1 - \mathbb{I}_i^{DC}\right) + c_{DC}^{site}\,\mathbb{I}_i^{DC}\right)x_i \le B \tag{6}$$

where $c_p^{cap}$ and $c_p^{op}$ denote the daily-equivalent capital and operating costs of a charger of type $p$, and $c_{AC}^{site}$, $c_{DC}^{site}$ represent the fixed daily site costs for AC and DC stations, respectively. $\mathbb{I}_i^{DC}$ is a binary variable equal to 1 if station $i$ contains at least one DC charger, and 0 otherwise.

### C. *Utilization-Driven Deployment Refinement*

The initial deployment provides spatial coverage but does not account for realized operational performance under simulated charging behavior. To incorporate simulation-based feedback into infrastructure adjustment, we propose a utilization-driven refinement algorithm. Based on energy-normalized utilization derived from BEAM outputs, charger quantities are iteratively increased for over-utilized types and reduced for under-utilized types according to predefined thresholds. When the total number of chargers at a station falls below the minimum requirement, the station is removed. Station locations are otherwise fixed during refinement. The overall procedure is summarized in Table I and terminates when the deployment remains unchanged over two consecutive iterations.

To account for heterogeneous charger power levels, we define charger utilization based on delivered energy normalized by theoretical charging capacity. For station $i$ and charger type $p$, the observed total charging energy is:

$$E_{i,p} = \sum_{s\in\mathcal{S}_{i,p}} e_s \tag{7}$$

where $e_s$ denotes the energy delivered during charging session $s$. The theoretical daily charging capacity is defined as:

$$C_{i,p} = n_{i,p} P_p T \tag{8}$$

where $n_{i,p}$ is the number of chargers of type $p$, $P_p$ is the rated charging power, and $T$ denotes the length of the evaluation horizon. The energy-normalized utilization is then given by:

$$u_{i,p} = \frac{E_{i,p}}{C_{i,p}} \tag{9}$$

This metric measures the fraction of theoretical charging capacity that is effectively utilized and enables direct comparison across charger power levels.

$$F_{i,p} = \int_0^T 1\{\text{all } n_{i,p} \text{ chargers busy}\}dt \tag{10}$$

The full-load hours (FLH) of charger type $p$ at station $i$ are defined as the total duration during which all $n_{i,p}$ chargers are simultaneously occupied.

## III. CASE STUDY

This study is based on the Multimodal Mobility Model (M3), an advanced activity agent-based modeling framework developed by the Network Modelling and Planning Group at Monash University. M3 integrates an activity-based demand model built upon ActivitySim [20] with a multi-modal agent-based dynamic traffic simulation implemented in MATSim. The two components are coupled through automated parameter calibration and model optimization procedures,

TABLE I. RULE PSEUDOCODE

**Algorithm: Utilization-driven rule-based deployment refinement**

**Input:** Initial deployment $C^{(0)} = \{n_{i,p}^{(0)}\}$, station set $I$ power set $P$; thresholds $u^{+}, u^{-}$; bounds $N_{min}, N_{max}$; stopping parameter $K$(max consecutive no-change iterations).

**Output:** Refined deployment $C^{*}$and objective $f(C^{*})$.

Set $t \leftarrow 0$, $k \leftarrow 0$.

repeat

  Run BEAM under deployment $C^{(t)}$and aggregate charging sessions over horizon $T$.

  For each $(i, p)$, compute delivered energy $E_{i,p} = \sum_{s \in \mathcal{S}_{i,p}} e_s$ and utilization $u_{i,p} = \frac{E_{i,p}}{n_{i,p}^{(t)} p T}$ .

  Set $C' \leftarrow C^{(t)}$.

  for each station $i \in I$do

  Compute total chargers $N_i = \sum_{p \in P} n'_{i,p}$.

    for each power $p \in P$do

      if $F_{i,p}^{(t)} \geq u^{+}$and $N_i < N_{max}$then

        $n'_{i,p} \leftarrow n'_{j,p} + 1$; $N_i \leftarrow N_i + 1$.

      else if $u_{i,p}^{(t)} < u^{-}$and $n'_{i,p} > N_{min}$then

        $n'_{i,p} \leftarrow n'_{i,p} - 1$; $N_i \leftarrow N_i - 1$.

      end if

    end for

    if $N_i < N_{\text{station}}^{min}$then remove station $i$(set $n'_{i,p} \leftarrow 0 \ \forall p$).

  end for

  if $C' = C^{(t)}$then $k \leftarrow k + 1$else $k \leftarrow 0$.

  Set $C^{(t+1)} \leftarrow C'$; $t \leftarrow t + 1$.

until $k \geq K$.

Run BEAM once more under $C^{(t)}$and compute $f(C^{(t)})$.

Return $C^{*} = C^{(t)}$and $f(C^{*})$.

enabling consistent representation of daily activity scheduling and network-level traffic dynamics.

The synthetic population used in this study represents the Greater Melbourne region (Australia). It consists of individuals with unique demographic identifiers and is generated through a population synthesis process that combines aggregated Census data, household attributes, and journey-to-work information. Control totals at the regional level are applied to ensure that the synthetic population matches observed demographic distributions and household totals. Using PopSim within ActivitySim, a virtual population is created, and each agent is assigned a complete daily activity plan. These activity schedules are derived from the Australian Census and the Victorian Integrated Survey of Travel and Activity (VISTA) [21].

For the present case study, the M3 dataset comprises 71,178 households among which 56,661 households with private vehicles are used in the simulation. An EV penetration rate of 5% is applied, resulting in 5,109 electric vehicles among 102,182 passenger vehicles. Furthermore, 60% of EV-owning households are assumed to have access to a Level 2 (7.2 kW) home charger, consistent with existing empirical assumptions in the literature [22].

TABLE II. PARAMETERS AND INFRASTRUCTURE COST ASSUMPTIONS

| Parameters | Value |
|---|---|
| Battery capacity (kWh) | 64 |
| Energy consumption rate (kWh/km) | 0.16 |
| 7.2/22/150 kW Charging margin (AUD/kWh) | 0.10/0.10/0.25 |
| 7.2/22/150 kW Charger CAPEX (AUD/day) | 1.10/1.90/41.1 |
| 7.2/22/150 kW Charger OPEX (AUD/day) | 0.5/0.8/12 |
| AC/DC Station cost (AUD/day) | 5.48/27.40 |
| Value of time (AUD/hour) | 30 |
| Vehicle operating cost (AUD/km) | 0.4 |

Table II summarises the key model parameters. All cost components are converted to daily equivalents to align with the BEAM simulation horizon. Capital expenditures are annualised assuming a ten-year lifetime, consistent with recent infrastructure cost assessments reported by the International Energy Agency [1]. Electricity tariffs are calibrated based on publicly reported charging prices in Australia. The charging margin is calculated as the difference between the retail charging tariff and the underlying electricity supply cost, and is used to represent operator revenue per kWh. The total investment budget is set to the level required to fully cover destination charging demand under the CMCLP deployment, corresponding to 2,600 in daily equivalent terms.

In the utilization-driven refinement stage, conservative thresholds are adopted to prevent excessive capacity adjustment, including a 5% minimum utilization rate and 2 full-load hours per day. These values are chosen within typical observed ranges and are not tuned for performance optimization.

Fig. 3 illustrates the modeling workflow. Panel (a) presents the spatial distribution of synthetic population activity locations.Panels (b)–(d) provide an illustrative example under the combined charging regime with utilization-driven refinement. Specifically, (b) shows representative EV trajectories and the resulting latent public charging demand,

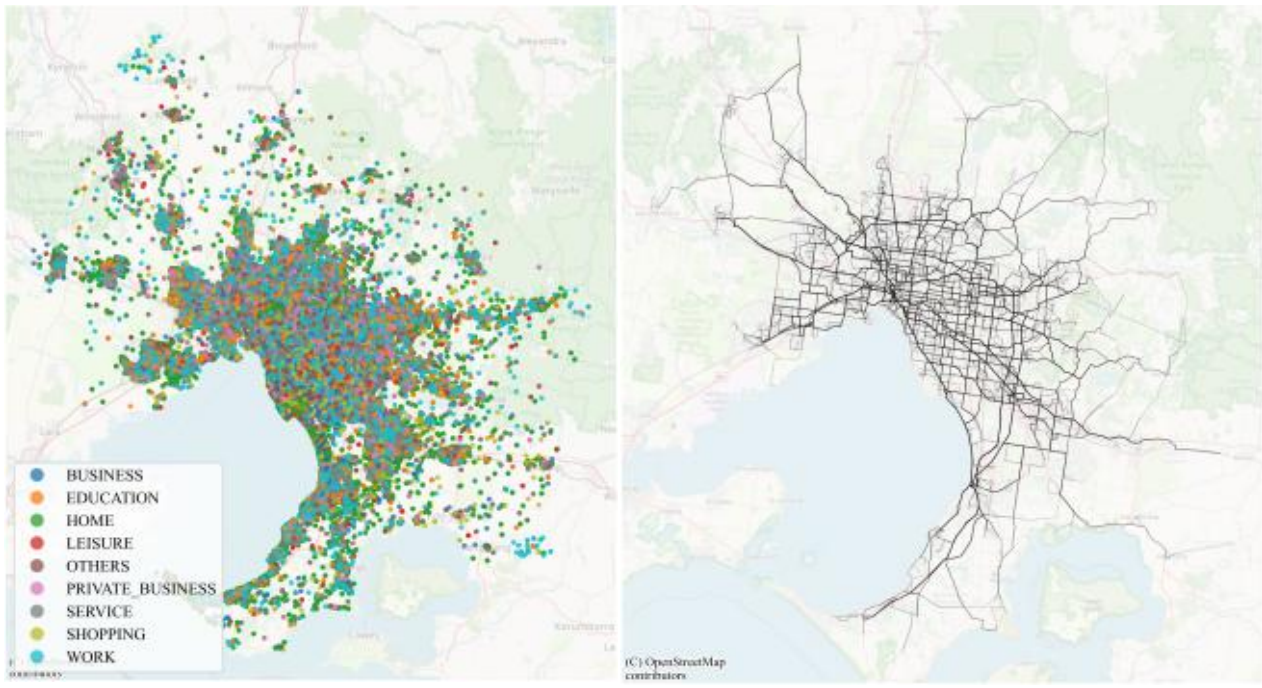

(a) Spatial Distribution of Synthetic Population Activities (b) Representative EV Trajectories

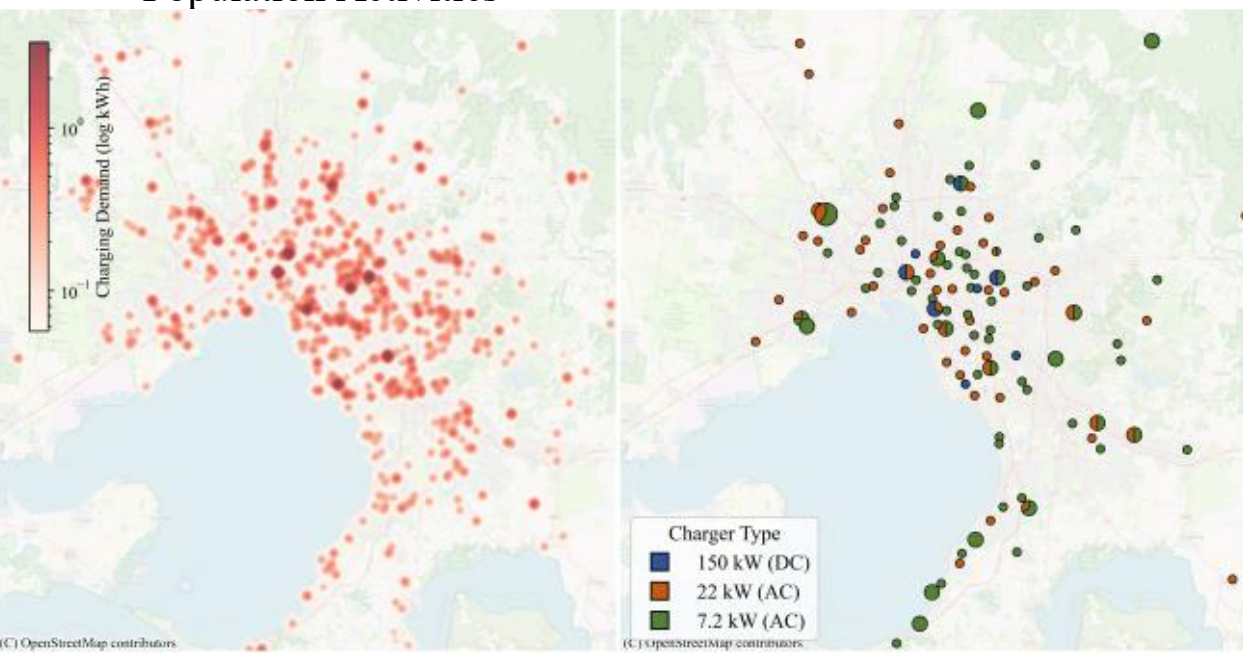

(c) Spatial Intensity of Latent Public Charging Demand (d) Charging Infrastructure Layout by Charger Type

Fig. 3. Synthetic Population Activity Distribution and EV Travel Patterns in Greater Melbourne

(c) visualizes the aggregated spatial intensity of charging demand, and (d) displays the corresponding refined infrastructure layout differentiated by charger type.

## IV. RESULT ANAYLSIS

To facilitate comparison, we use a compact notation to distinguish deployment types and refinement stages. The suffixes D, E, and C represent destination, en-route, and combined charging deployments, respectively. CMCLP-* denotes layouts directly derived from the CMCLP formulation, while REF-* refers to the corresponding layouts after utilization-driven refinement.

### A. *Charging Infrastructure Deployment*

As shown in Fig. 3(d), AC chargers are widely dispersed across the metropolitan area, consistent with destination-based charging near activity locations. In contrast, DC fast chargers are concentrated in central urban areas and major demand corridors, where traffic density is highest. This spatial configuration corresponds to their functional differentiation, with AC chargers primarily serving longer dwell-time activities that require broader coverage, and DC fast chargers supporting en-route and high-throughput demand in high-traffic areas.

Table III summarizes the infrastructure composition under different deployment scenarios. In all three cases, utilization refinement reduces the total number of stations and chargers, indicating more efficient capacity allocation based on observed usage. Compared with the CMCLP-based layouts, the refined configurations contain fewer stations and a more selective allocation of chargers, with a substantially larger reduction in AC slow chargers than in DC fast chargers. This reflects the relatively dispersed and stochastic nature of destination charging demand, which results in lower utilization rates and allows underperforming facilities to be removed with limited impact on overall system-wide charging activity.

TABLE III. CHARGING INFRASTRUCTURE COMPOSITION ACROSS DEPLOYMENT SCENARIO

| **Layout** | **Station count** | | **Charger count** | | |
|---|---|---|---|---|---|
| | ***DC*** | ***AC*** | ***7.2 kW*** | ***22 kW*** | ***150 kW*** |
| CMCLP-D | / | 248 | 320 | 258 | / |
| REF-D | | 150 | 233 | 84 | / |
| CMCLP-E | 19 | / | / | / | 39 |
| REF-E | 19 | / | / | / | 36 |
| CMCLP-C | 8 | 146 | 178 | 164 | 16 |
| REF-C | 8 | 105 | 131 | 76 | 16 |

### B. *System Performance*

Table IV reports system-level performance metrics for different charging infrastructure deployments.

On the operator side, charging revenue varies across deployment types. The destination charging deployment yields the lowest revenue, while the en-route charging deployment produces the highest revenue, reflecting the higher unit energy prices associated with DC fast charging. After refinement, charging revenue changes remain within 1% across all deployments, indicating that overall charging demand are largely preserved. Deployment costs are consistently reduced under refinement deployments. The largest cost reduction is observed under the destination charging deployment, consistent with the greater reductions in both station and charger counts.

On the user side, detour distance per vehicle is defined as the additional network travel incurred for each en-route charging event relative to the Euclidean origin–destination distance scaled by 1.4 to approximate road circuity [23]. Under the en-route deployment, refinement leads to only minor changes in average detour distance, consistent with limited infrastructure adjustments.

Notably, under the combined charging deployment, the average detour distance decreases substantially and becomes lower than that under the en-route charging deployment, even though the deployment of DC fast chargers remain unchanged. This suggests that refinement of the slow-charging layout enhances the overall spatial coordination between charging demand and infrastructure, thereby indirectly improving en-route charging efficiency. The underlying mechanism is further discussed in the following subsection.

Across all scenarios, fewer than 4% of EVs experience negative SoC (NSoC) events during the simulation day. The highest incidence occurs under the destination charging deployment, and this count increases slightly after refinement, reflecting reduced redundancy in charging capacity. In contrast, both the en-route and combined deployments exhibit lower NSoC counts, which further decline marginally after refinement, indicating improved energy feasibility when fast-charging provision is better aligned with demand.

Fig. 4 compares system-level performance across the charging deployments using three metrics. Operator net benefit is defined as charging revenue minus deployment costs. User generalized cost includes detour costs and penalties for NSoC events, with a 100 AUD penalty per affected vehicle. System total cost reflects a social objective that excludes charging revenue and aggregates deployment costs and user generalized cost.

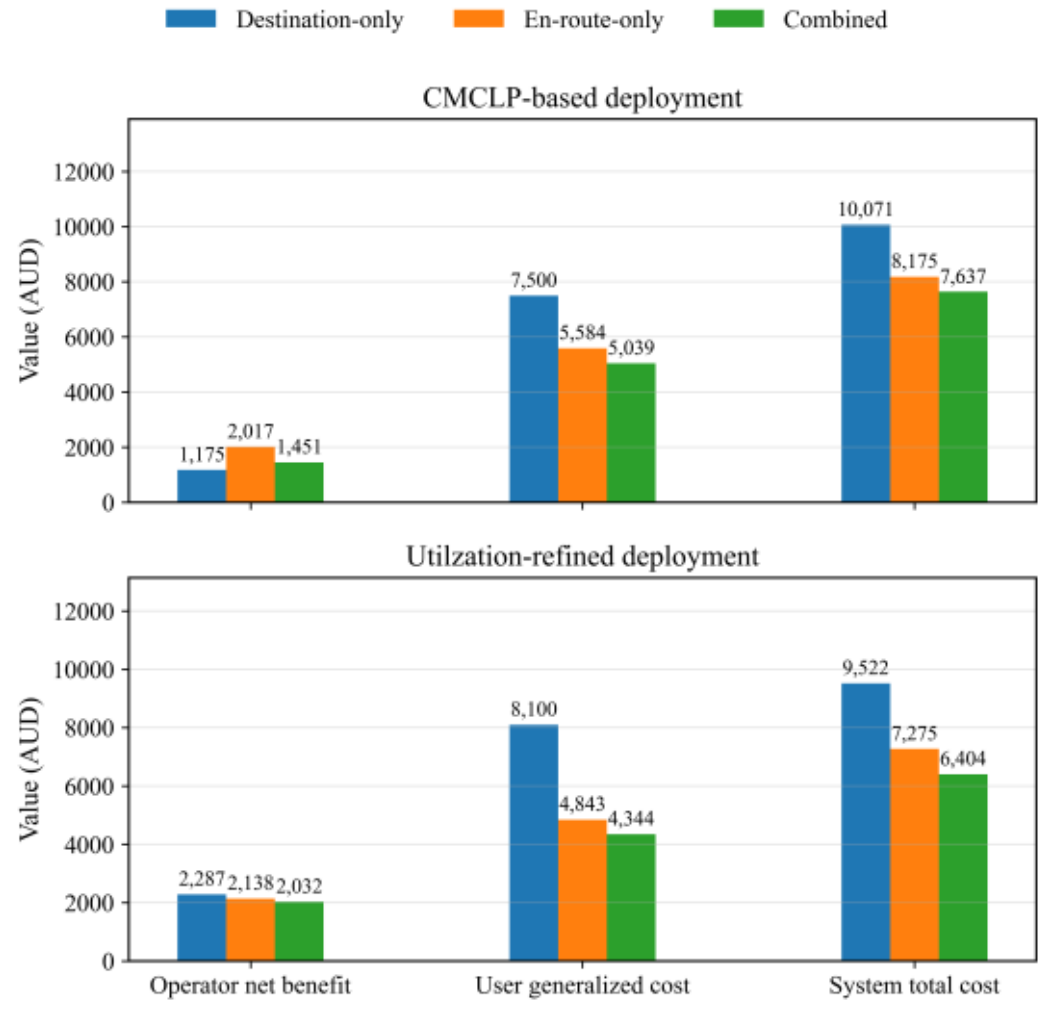

Fig. 4. System-Level Performance Comparison across Charging Deployments

TABLE IV. SYSTEM-LEVEL PERFORMANCE COMPARISON ACROSS DEPLOYMENT LAYOUTS AND CHARGING REGIMES

| **Layout** | **Operator side** | | **User side** | | | |
|---|---|---|---|---|---|---|
| | ***Charging revenue (AUD)*** | ***Deployment cost (AUD)*** | ***Detour distance per vehicle (km)*** | ***Normalized Detour Ratio*** | ***Detour cost (AUD)*** | ***Negative SoC counts*** |
| CMCLP-D | 3745.98 | 2570.78 | / | / | / | 75 |
| REF-D | 3709.06 (-0.99%) | 1422.06 (-44.68%) | / | / | / | 81 (+7.96%) |
| CMCLP-E | 4608.29 | 2591.20 | 5.99 | 1.14 | 483.70 | 51 |
| REF-E | 4570.24 (-0.83%) | 2432.00 (-6.15%) | 5.48 | 1.11 | 442.51 (-8.52%) | 44 (-14.72%) |
| CMCLP-C | 4049.54 | 2598.50 | 11.13 | 1.21 | 638.58 | 44 |
| REF-C | 4091.48 (+1.04%) | 2059.67 (-20.74%) | 6.00 | 1.13 | 344.25 (-46.09%) | 40 (-7.57%) |

### C. User Charging Behavior

To further examine user behavioral changes induced by refinement, we compare charging start time and duration distributions using CMCLP-C and REF-C layout, as shown in Fig. 5.

For charging start time, en-route charging shows a 6.10% reduction in standard deviation, while the mean remains nearly unchanged, indicating a slightly more concentrated temporal distribution. In contrast, destination charging start time exhibits a 1.49% increase in mean along with a modest increase in dispersion. The upward shift in mean suggests that, with fewer but more intensively utilized slow chargers, some charging sessions are initiated later in the day. This temporal adjustment may alleviate simultaneous demand during peak periods and increase fast charger availability when en-route charging is triggered, contributing to shorter detour distances.

Charging duration remains stable for en-route charging, with an average of approximately 20 minutes in both scenarios. Destination charging durations average around 2 hours and include longer sessions associated with extended activities. After refinement, the mean destination charging duration increases by 5.73%, consistent with the slight delay in charging initiation and reflecting more sustained slow charger utilization.

Overall, these results indicate that refinement subtly reshapes the temporal and utilization structure of charging activities, contributing to improved infrastructure efficiency and reduced detour costs. In particular, a more appropriate allocation of AC slow chargers enhances destination charging availability, which in turn influences en-route charging behavior and further improves user-level benefits.

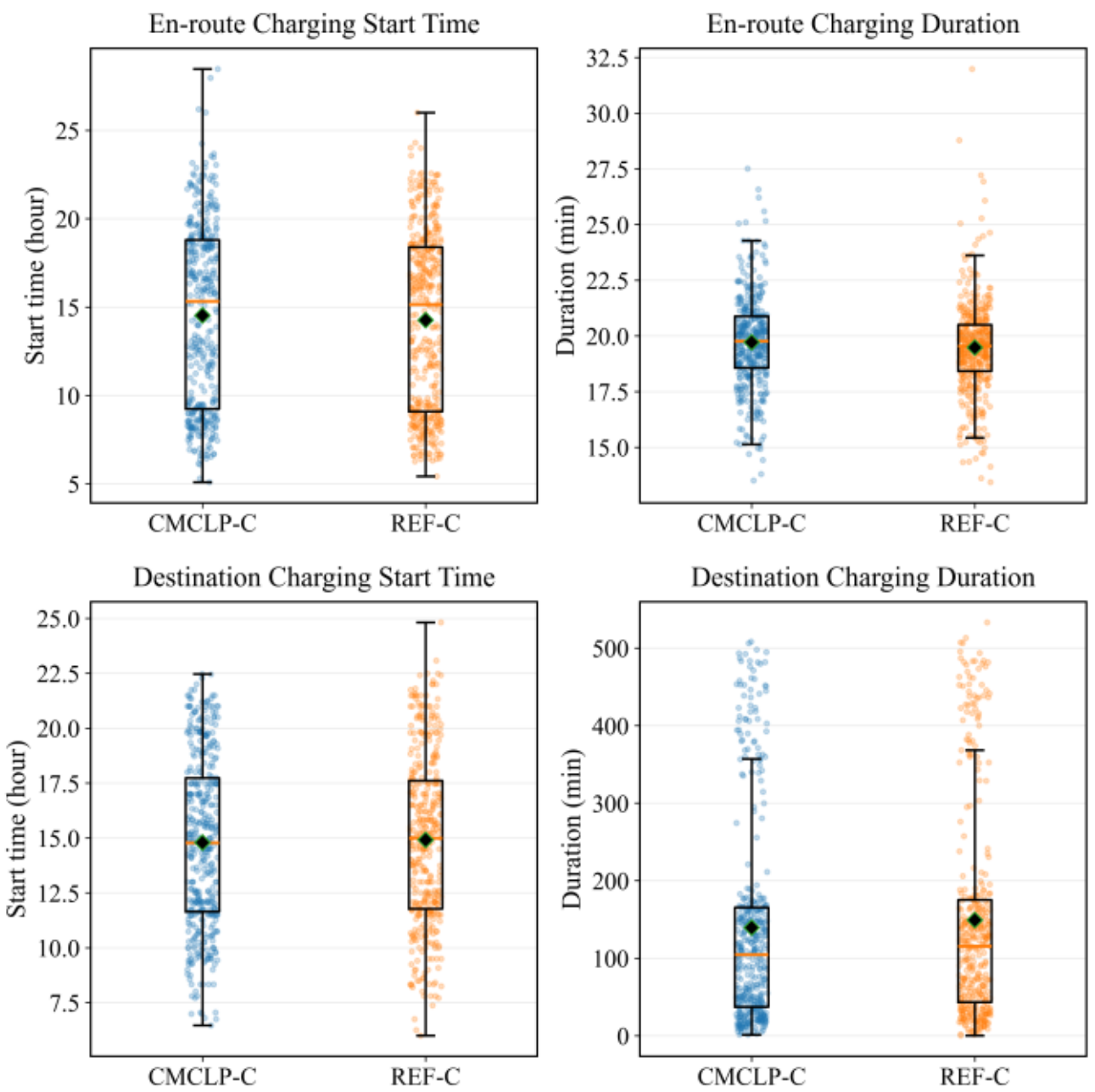

Fig. 5. Distributions of Charging Start Time and Duration for En-route and Destination Charging under CMCLP-C and REF-C

## V. CONCLUSION

This study employs an agent-based model to generate latent charging demand under three charging regimes and evaluates system-level performance across two deployments strategies: a CMCLP-based layout and a utilization-driven refined layout. In total, six deployment configurations are analyzed to assess how infrastructure design shapes charging behavior and overall system outcomes.

Across all charging regimes, utilization-driven refinement reduces total system cost, with the largest reduction observed under the combined regime. Under the destination charging regime, refinement yields the greatest decrease in deployment cost, but this is accompanied by a slight increase in user generalized cost, indicating a trade-off between infrastructure efficiency and user-level performance.

Under the combined charging regime, adjustments in AC slow charger allocation modify destination charging patterns, which subsequently influence en-route charging behavior. Specifically, a more concentrated slow charger deployment is associated with modest delays and extended destination charging durations, corresponding to reduce en-route detour distance.

Overall, the findings underscore the importance of jointly considering en-route charging and destination charging regimes in infrastructure planning and demonstrate how utilization-driven refinement reshapes both system cost structure and behavioral outcomes. This study focuses on illustrating behavioral and system-level implications rather than introducing a new optimization algorithm. Future research will further examine alternative refinement criteria and extended optimization formulations to enhance system robustness and efficiency.